\documentclass[10pt,twocolumn,letterpaper]{article}

\usepackage{wacv}              

\usepackage{graphicx}
\usepackage{amsmath}
\usepackage{amssymb}
\usepackage{booktabs}
 \usepackage{multirow}
\usepackage{caption}
\usepackage{setspace}
\usepackage{xcolor}
\usepackage[pagebackref,breaklinks,colorlinks]{hyperref}
\usepackage[accsupp]{axessibility}
\newtheorem{definition}{Definition}
\newtheorem{lemma}{Lemma}
\usepackage[capitalize]{cleveref}
\crefname{section}{Sec.}{Secs.}
\Crefname{section}{Section}{Sections}
\Crefname{table}{Table}{Tables}
\crefname{table}{Tab.}{Tabs.}

\def\wacvPaperID{1487} 
\def\confName{WACV}
\def\confYear{2025}

\begin{document}

\title{$\mathcal{J}$-Invariant Volume Shuffle for Self-Supervised Cryo-Electron Tomogram Denoising on Single Noisy Volume}

\author{Xiwei Liu$^{1}$ \quad Mohamad Kassab$^{1}$ \quad Min Xu$^{1, 2}$ \quad Qirong Ho$^{1,*}$\\
$^{1}$Mohamed bin Zayed University of Artificial Intelligence \quad $^{2}$Carnegie Mellon University\\
{\tt\small \{Xiwei.Liu, Mohamad.Kassab, Min.Xu, Qirong.Ho\}@mbzuai.ac.ae}
}

\maketitle

\begin{abstract}
   Cryo-Electron Tomography (Cryo-ET) enables detailed 3D visualization of cellular structures in near-native states but suffers from low signal-to-noise ratio due to imaging constraints. Traditional denoising methods and supervised learning approaches often struggle with complex noise patterns and the lack of paired datasets. Self-supervised methods, which utilize noisy input itself as a target, have been studied; however, existing Cryo-ET self-supervised denoising methods face significant challenges due to losing information during training and the learned incomplete noise patterns. In this paper, we propose a novel self-supervised learning model that denoises Cryo-ET volumetric images using a single noisy volume. Our method features a U-shape $\mathcal{J}$-invariant blind spot network with sparse centrally masked convolutions, dilated channel attention blocks, and volume-unshuffle/shuffle technique. The volume-unshuffle/shuffle technique expands receptive fields and utilizes multi-scale representations, significantly improving noise reduction and structural preservation. Experimental results demonstrate that our approach achieves superior performance compared to existing methods, advancing Cryo-ET data processing for structural biology research. Code is available at \url{https://github.com/Xiwei-web/SelfCryoET}.
\end{abstract}
\section{Introduction}
\label{sec:intro}
\begin{table*}[t]
    \centering
    \captionsetup{skip=0pt} %
    \begin{spacing}{1.2}
    \caption{Comparison of the exiting self-supervised denoising methods for Cryo-ET}
     \label{tab:Comparison of the exiting methods}
    \end{spacing}
    \resizebox{\textwidth}{!}{
    \begin{tabular}{lcccc}
        \toprule
        Feature & \textbf{Ours} & SC-Net & NMSG & Noise2Void\\
        \midrule
        Main Architecture & \textbf{U-shape $\mathcal{J}$-invariant BSN } & Tradition U-Net & Tradition U-Net  & Tradition U-Net  \\
        
        Key Components & \textbf{CMC+DCA+VU/S} & Conv3D+BN3D+LReLU & Conv3D+BN3D+LReLU & Conv3D+BN3D+LReLU \\

        $\mathcal{J}$-invariance & \textbf{Meet} & Meet (information loss) & Doesn't Meet & Meet \\

        downsampling/upsampling & \textbf{volume-unshuffle/shuffle} & max pooling / interpolation  & max pooling / interpolation & max pooling / interpolation \\
        \bottomrule
    \end{tabular}
    }
    \vspace{-1em}
\end{table*}
\renewcommand{\thefootnote}{} 
\footnotetext{$^{*}$Corresponding author}
Cryo-Electron Tomography (Cryo-ET) is an advanced imaging technique that provides the visualization of cellular structures and macromolecular complexes in three dimensions at near-native states \cite{turk2020promise}. Combining the principles of Cryo-Electron Microscopy (cryo-EM) with tomographic reconstruction \cite{frank1987three}, Cryo-ET creates detailed 3D images of specimens that have been rapidly frozen, preserving their natural structure without the need for staining or chemical fixation. However, Cryo-ET images typically suffer from a low signal-to-noise ratio (SNR). Several factors contribute to this issue: low electron doses to prevent radiation damage result in weaker signals \cite{adrian1984cryo}, the thickness of biological specimens causes electron scattering, and the cryogenic conditions introduce additional noise. Consequently, the high levels of noise can obscure fine structural details, complicate the tomographic reconstruction process, and make data interpretation challenging \cite{bepler2020topaz}. 

Traditional denoising methods \cite{buades2005non},~\cite{frangakis2001noise},~\cite{tomasi1998bilateral},~\cite{munch2009stripe} often fall short in handling the complex noise patterns inherent in Cryo-ET data. These methods rely on predefined assumptions about the noise and signal characteristics, limiting their flexibility and performance. In recent years, deep neural networks (DNNs) have emerged as powerful tools for image denoising \cite{jain2008natural},~\cite{xie2012image},~\cite{zhang2017beyond}. Supervised learning approaches, which rely on large datasets of paired noisy and clean images, have shown promise. However, acquiring paired datasets for Cryo-ET is extremely challenging due to the difficulty in obtaining high-quality ground truth data.

To address the abovementioned challenges, self-supervised denoising methods have gained attention. Unlike supervised methods, the model in self-supervised learning, which uses the same image for both input and target, has to satisfy $\mathcal{J}$-invariance \cite{batson2019noise2self}. $\mathcal{J}$-invariance ensures that the prediction for each pixel is not influenced by its original value. Blind-spot networks (BSNs) \cite{krull2019noise2void}, which satisfy the $\mathcal{J}$-invariance requirement, have been successfully implemented for denoising noisy inputs in a self-supervised manner \cite{lee2022ap}\cite{zhang2023mm}. As summarized in Table \ref{tab:Comparison of the exiting methods}, SC-Net \cite{yang2021self} adapts a volumetric blind-spot strategy on input volume to maintain $\mathcal{J}$-invariance. Nevertheless, SC-Net only uses masked volumetric patches for model training, which can lead to resource wastage and loss of information during training. Both SC-Net and NMSG \cite{gulrajani2017improved} utilize U-Net as their primary architecture, while NMSG has to utilize Generative Adversarial Networks (GANs) \cite{gulrajani2017improved} to learn and model the noise patterns present in Cryo-ET data, generating pairs of noisier and noisy images for training. By constructing a BSN with centrally masked convolutions (CMC) and dilated convolution layers (DCL), we can address these limitations without the need for GANs to pre-learn Cryo-ET noisy patterns. Additionally, the centrally masked convolutions in BSN ensure that all voxels participate in training \cite{zhang2023unleashing}, unlike the mask-in-input strategy of SC-Net that leads to information loss. However, most current methods use BSN built with stacked CMC and DCLs instead of U-shape network \cite{wu2020unpaired},~\cite{lee2022ap},~\cite{zhang2023mm}, as traditional downsampling operation breaks $\mathcal{J}$-invariance.

As we know that U-Net structures are particularly effective in capturing long-range dependencies and coarse-to-fine representations \cite{cho2021rethinking},~\cite{yue2020dual},~\cite{zamir2021multi}. To keep leveraging U-Net in self-supervised learning, we propose volume-unshuffle/shuffle (VU/S), a suitable downsampling/upsampling technique for volumetric images that preserves $\mathcal{J}$-invariance. We incorporated sparse centrally masked convolutions, dilated channel attention (DCA) block \cite{chen2022simple} and volume-unshuffle/suffle to build a novel $\mathcal{J}$-invariant U-shape self-supervised learning model, specifically designed for denoising Cryo-ET volumetric images using a single noisy volume. Our method is validated through extensive experiments on both simulated and real-world Cryo-ET datasets. The results demonstrate that our approach outperforms existing self-supervised denoising methods based on single noisy volumes, achieving superior noise reduction and structural preservation.

Our contribution can be summarized as follows:

    $\bullet$ A novel U-shape self-supervised volumetric image denoising model is introduced, by adapting sparse centrally masked convolution and dilated channel attention (DCA) blocks to build a $\mathcal{J}$-invariant blind spot network.
    
    $\bullet$ We propose a novel downsampling/upsampling technique called volume-unshuffle/shuffle, which preserves the $\mathcal{J}$-invariance property essential for effective denoising. This technique expands the receptive fields and utilizes multi-scale representation, significantly improving the model’s ability to capture long-range dependencies and structural details within the Cryo-ET volumetric data. 
    
    $\bullet$ We introduce dilated channel attention (DCA) block to the domain of Cryo-ET volumetric images denoising, which effectively incorporate global context through channel attention. 
    
    $\bullet$ We design a combined loss function that balances noise suppression with structure preservation. This includes structural reconstruction loss, contrast guidance loss, edge enhancement loss, and total variation loss, ensuring that the denoised output retains essential structural details.

\begin{figure*}[h!]
    \centering
    \includegraphics[width=0.85\textwidth]{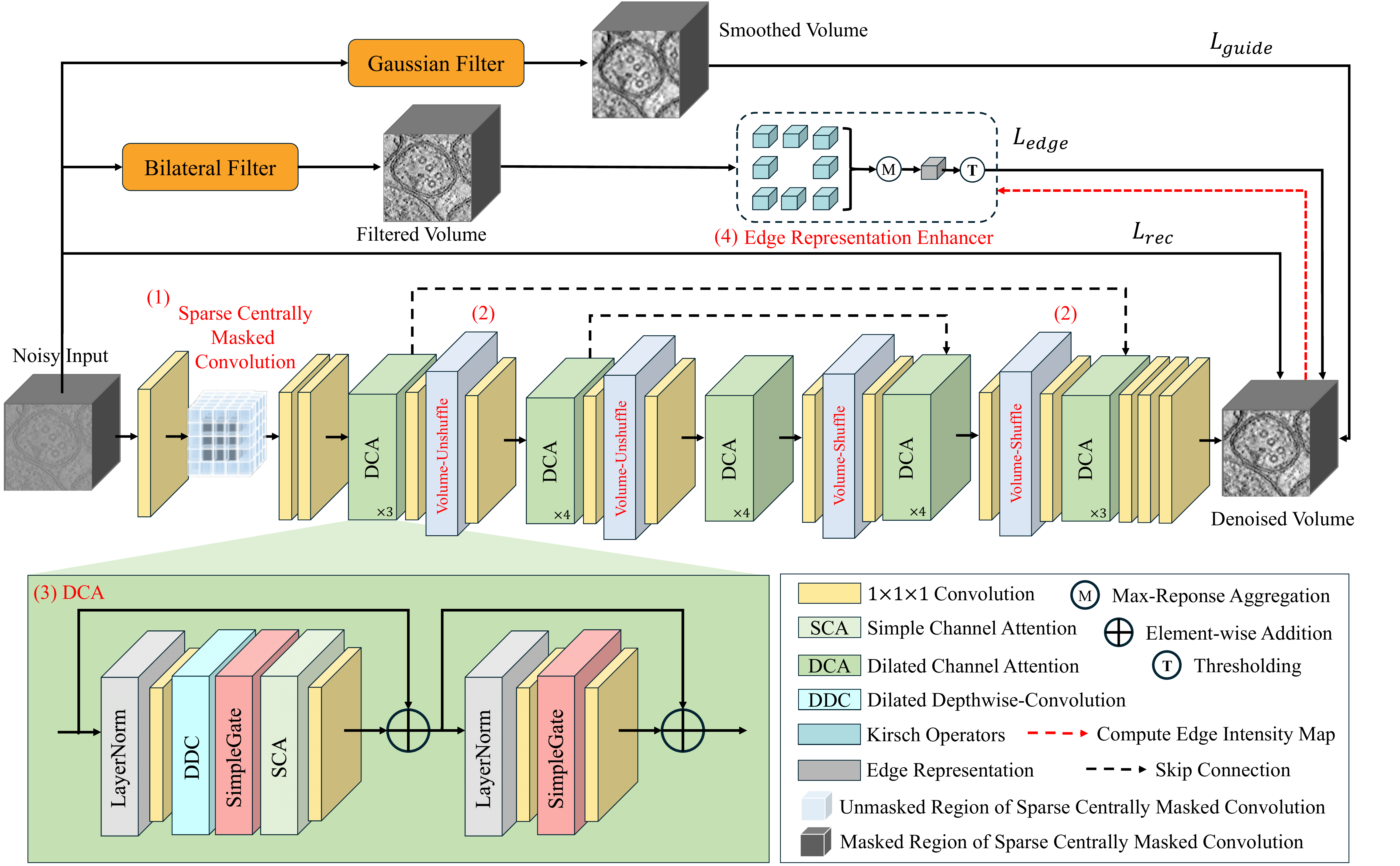}
    \captionsetup{skip=0pt} %
    \begin{spacing}{0.5}
    \caption{The architecture of proposed method. The noisy input volume is first preprocessed using (i) 3D Gaussian filters to generate a smoothed volume and (ii) bilateral filters to create a edge preserved filtered volume, respectively. We construct the U-shape BSN with 4 main contributions: (1) Sparse centrally masked convolution, (2) Volume Unshuffle/Shuffle and (3) DCA blocks, with (4) an edge representation enhancer utilizes the filtered volume to guide model training. Refer to Section~\ref{sec:main_components} for details.}
    \label{fig: workflow overview}
    \end{spacing}
    \vspace{-1em}
\end{figure*}

\section{Related Works}

\textbf{Traditional volumetric image denoising.} 
Cryo-ET data processing traditionally involves denoising either before or after tomographic reconstruction. Methods that denoise projections before reconstruction \cite{maiorca2012improving} often fail to maintain the 2D-to-3D relationship defined by the Fourier-slice theorem \cite{ng2005fourier}, resulting in overly smoothed volumes. Linear filters such as low-pass filters, Gaussian filters, and dose filtering are easy to adapt and tune but often lack the sophistication needed for complex noise patterns. Nonlinear filters include the iterative median filter \cite{van2007efficient}, nonlinear anisotropic diffusion (NAD) \cite{frangakis2001noise}, bilateral and trilateral filters \cite{tomasi1998bilateral}, Non-local Means (NLM) \cite{buades2005non}, and BM4D (Block-Matching and 4D Filtering) \cite{maggioni2012nonlocal}. These traditional methods often rely on predefined assumptions about the data, limiting their flexibility and performance in varying scenarios.

\textbf{Supervised learning image restoration.} DNNs have shown great promise in image denoising. Supervised learning approaches, which rely on large datasets of paired noisy and clean images, have demonstrated significant advancements. Jain et al. \cite{jain2008natural} first employed DNN for image denoising, achieving results comparable to the best traditional algorithms at the time. Subsequently, more DNN-based methods were proposed \cite{zhang2017beyond}\cite{zhang2018ffdnet}, \cite{xie2012image}, \cite{dong2018denoising}, \cite{agostinelli2013adaptive}. Encoder-Decoder networks \cite{mao2016image} and GANs \cite{isola2017image} have also been used for image restoration. However, acquiring high-quality paired datasets for Cryo-ET is extremely challenging, limiting the application of supervised learning methods in this field.

\textbf{Self-supervised learning image restoration.} Self-supervised denoising methods do not require paired noisy and clean datasets, making them particularly suitable for Cryo-ET data. Noise2Noise (N2N) \cite{lehtinen2018noise2noise} was the first self-supervised algorithm that achieved performance comparable to supervised image denoising by training on only aligned noisy-noisy image pairs. Topaz-Denoiser \cite{bepler2020topaz} and CryoCARE \cite{buchholz2019content}, based on the N2N concept, have been effectively applied to Cryo-EM and Cryo-ET. These approaches train a neural network using pairs of noisy images to distinguish between signal and noise without needing clean reference images. However, acquiring a large number of noisy volumes of the same sample can be impractical and time-consuming, limiting the availability of training data. Noisier2Noise \cite{moran2020noisier2noise} extended this concept by adding synthetic noise to noisy images to create training pairs. The NMSG \cite{yang2024self} method leverages GANs to pre-learn the noise pattern in tomogram, generating pairs of noisier and noisy images for training. However, NMSG extracts several 2D noise slices only from the top and bottom slices in the raw noisy volume as the learning targets for noise modeling. This approach may not capture the full variability and intricacies of the noise present throughout the entire 3D volume, due to the inherent complexity and heterogeneity of noise patterns in Cryo-ET data. Derived from Noise2Void \cite{krull2019noise2void}, SC-Net \cite{yang2021self} proposed a volumetric blind-spot replacement strategy, designed for Cryo-ET, extending the blind-spot concept into three dimensions. However, only the masked volumetric patches participate in model training each time, leading to wasted resources and time. Additionally, the masked volumetric patches no longer appear in the training process, resulting in the loss of information during training.

As the most widely used self-supervised denoising method, BSN was firstly proposed in Noise2Void, which is a special CNN that masks pixel in the center of the receptive field, and uses the surrounding information to reconstruct the information of the masked pixels. Variations like Laine19 \cite{laine2019high}, D-BSN \cite{wu2020unpaired}, AP-BSN \cite{lee2022ap}, MM-BSN \cite{zhang2023mm} and PUCA \cite{jang2024puca} have prove their effectiveness of learning and removing the complicated nosie pattern. These advancements highlight the promising future to refine self-supervised denoising techniques for Cryo-ET, addressing the unique challenges posed by the complex noise patterns and the lack of paired datasets.

\section{Preliminaries}
Following the assumption proposed in the work by Yang \cite{yang2021self}, we assume that a projection image $I_n(x, y)$ in cryo-ET represents a discrete observation of the projection $P_n(x, y)$ with additive Gaussian noise $N(x, y)$. This can be expressed as $I_n = P_n + N_n$. Therefore, we have:

\begin{lemma}
    The additive Gaussian noise in 2D projection remains Gaussian noise in the 3D reconstruction.
\end{lemma}
\begin{equation}
    V(x) = \Psi(x) + N(x),
\end{equation}
where $V(x)$ represents the volumetric image reconstructed from a series of $I_n$, $\Psi(x)$ is the ideally noise-free image reconstructed from $P_n$, and $N(x)$ is the noise in the 3D space. This assumption provides the theoretical foundation for our method, ensuring that the noise characteristics are consistent during the transition from 2D to 3D.

$\mathcal{J}$-invariance is a critical concept in self-supervised image denoising, aimed at preventing models from learning to replicate noisy inputs directly. Noise2Self \cite{batson2019noise2self} firstly point out that the self-supervied denoising function $f$ should be $\mathcal{J}$-invariant, the definition is as follows.

\begin{definition}
Consider a partition $\mathcal{J}$ of the dimensions $\{1, \ldots, m\}$, and select $J$ as a subset from $\mathcal{J}$. A function $f : \mathbb{R}^m \to \mathbb{R}^m$ is defined as $\mathcal{J}$-invariant if, for any given partition $J$, the output $f(x)_J$ is entirely independent of the input values $x_J$ within that subset. In other words, $f$ maintains $\mathcal{J}$-invariance if $f(x)_J$ remains constant regardless of changes in $x_J$, where $f(x)_J$ and $x_J$ denote the values of $f(x)$ and $x$ restricted to $J$, respectively. This ensures that the prediction for each pixel relies solely on the surrounding pixel values, not on the pixel itself.
\end{definition}

\section{Methodology}
\subsection{Network Architecture}
The proposed model begins with the raw noisy volumetric Cryo-ET data as shown in Figure \ref{fig: workflow overview}. Initially, the data undergoes preprocessing using Gaussian and bilateral filters to generate smoothed and edge-preserved filtered volumes, respectively. These volumes are then divided into smaller overlapping patches. The patches are fed into a 2-level U-shape encoder-decoder network \cite{ronneberger2015u} designed for volumetric image denoising.

The encoding path starts with a 5$\times$5$\times$5 sparse centrally masked convolution layer, which ensures $\mathcal{J}$-invariance by preventing the output voxels from being influenced by their corresponding input voxels. This layer is followed by volume-unshuffle operations at each level to downsample the input, effectively increasing the channel dimensions while reducing the spatial dimensions, capturing multi-scale features and expanding the receptive fields. Within the encoder, features are processed by DCA blocks. These blocks with the channel attention mechanisms integrate global context, enhancing the network's ability to distinguish between noise and meaningful features.

The decoding path restores the spatial resolution using volume-shuffle operations, which reverse the downsampling process of volume-unshuffle. Skip connections are used to merge features from corresponding encoder levels, ensuring that detailed structures and textures are retained in the restored images. The final output image is generated by sequentially applying 1$\times$1$\times$1 convolution layers to the merged features, resulting in a denoised image.

\subsection{Main Components}
\label{sec:main_components}

\textbf{(1) Sparse Centrally Masked Convolution.}
In Cryo-ET data, the complexity of three-dimensional structures and significant noise levels necessitate a robust approach for effective denoising. Traditional centrally masked convolutions, which mask only the central pixel, are inadequate for this task. Instead, we employ a sparse centrally masked convolution, using a 5$\times$5$\times$5 kernel with the central 3$\times$3$\times$3 region entirely masked. This design offers several advantages. First, the larger receptive field captures broader contextual information, crucial for understanding volumetric data and differentiating between signal and noise. Second, by masking a larger central region, we ensure $\mathcal{J}$-invariance, preventing the model from learning identity mappings and enhancing its generalization to unseen noisy data. Third, the method adapts to the strong coherent ultrastructures in Cryo-ET data, balancing noise reduction and structure preservation. This approach focuses on the surrounding context, minimizing the influence of noise and preserving fine structural details, ultimately improving denoising performance.

\textbf{(2) Volume Unshuffle/Shuffle.}
Inspired by~\cite{jang2024puca}, we propose a volume unshuffle/shuffle technique that significantly increases the receptive fields of the neural network, thereby enhancing its ability to capture long-range dependencies and multi-scale context within volumetric images. Downsampling, achieved through this technique, helps the denoiser understand the relationships and dependencies between different parts of the volume more effectively. In 2D image processing, pixel-unshuffle/shuffle methods that preserve original pixels have proven to be effective in image restoration. However, directly applying pixel-unshuffle to 3D volumetric images with voxels disrupts the $\mathcal{J}$-invariance property of BSN, which is essential for preventing the network from learning identity mappings (shown in supplymentary material S1).

\vspace{-1.0em}
\begin{figure}[h]
    \centering
    \includegraphics[width=0.48\textwidth]{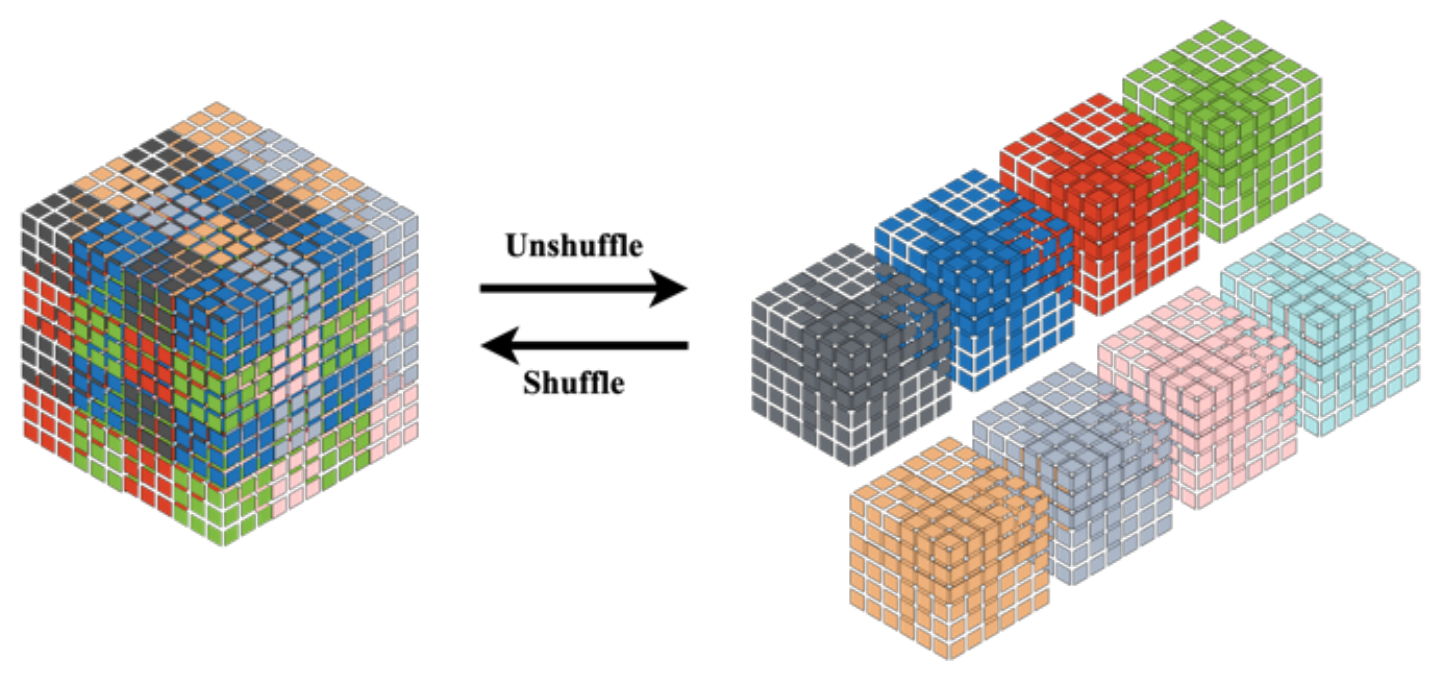} 
    \captionsetup{skip=0pt} %
    \begin{spacing}{0.5}
    \caption{Volume unshuffle and shuffle.}
    \label{fig:Volume shuffle and unshuffle}
    \end{spacing}
\end{figure}

To address this issue, we introduce the volume-unshuffle technique for volumetric tomogram images. Volume-unshuffle transforms a tensor of size \( D \times H \times W \times C \) into a reshaped tensor of size \( \frac{D}{v} \times \frac{H}{v} \times \frac{W}{v} \times (C \cdot v^3) \), where $v$ represents the shuffle volume size, as illustrated in Figure \ref{fig:Volume shuffle and unshuffle}. This transformation process ensures that the spatial information is redistributed into the channel dimension, thereby preserving the spatial structure while expanding the receptive field. This process can be mathematically defined as follows:

\vspace{-1.0em}
\small
\begin{equation}
    \text{Volume-Unshuffle}(y^{(l)}_{i,j,k,m}, v) = y^{(l)}_{i',j',k',m'}
\end{equation}
\begin{equation}
    i' =  v \left\lfloor \frac{i}{v^3} \right\rfloor + (i\!\!\!\!\mod v)
\end{equation}
\begin{equation}
    j' = v\left\lfloor \frac{j}{v^3} \right\rfloor + (j\!\!\!\!\mod v)
\end{equation}
\begin{equation}
    k' = v \left\lfloor \frac{k}{v^3} \right\rfloor + (k\!\!\!\!\mod v)
\end{equation}
\vspace{-1em}
\begin{equation}
    m' =  C (v \left\lfloor \frac{i \!\!\!\!\mod v^3}{v} \right\rfloor + \left\lfloor \frac{j\!\!\!\! \mod v^3}{v} \right\rfloor + \left\lfloor \frac{k\!\!\!\! \mod v^3}{v} \right\rfloor ) + m
\end{equation}
\normalsize
As demonstrated in Supplementary Material S2, our networks ensure the preservation of $\mathcal{J}$-invariance by employing volume-unshuffle with a volume size \( v=3 \) equals to dilation factor \( d = 3 \). This integration of volume-unshuffle in BSN allows for effective downsampling operations while maintaining the independence of each voxel's prediction from its input value. Moreover, the volume-shuffle operation, which acts as the reverse process of volume-unshuffle, facilitates the upsampling of downsampled feature maps, restoring them to their original volume sizes. This operation ensures that detailed structural information is preserved during the denoising process, leading to superior performance in noise reduction and ultrastructure maintenance.

\textbf{(3) Dilated Channel Attention (DCA).}
The DCA blocks in the proposed network work in conjunction with the sparse centrally masked convolution at the beginning of the network to form the BSN. To fulfill the blind-spot characteristic, we incorporate a 3-dilated 3$\times$3$\times$3 depth-wise convolution (DDC) before the gating and attention mechanisms. This follows D-BSN \cite{wu2020unpaired}, which uses a centrally masked convolution kernel size of \(k = 2d - 1\). The DCA blocks leverage this structure to incorporate global context via attention mechanisms, enhancing the denoising process by focusing on relevant features and suppressing noise.

The DCA block consists of several components: LayerNorm \cite{ba2016layer}, a 1$\times$1$\times$1 convolution, skip connection, SimpleGate \cite{he2016deep} , and Simplified Channel Attention (SCA) \cite{chen2022simple}, along with DDC. LayerNorm is used to normalize the input features, ensuring stable training. The 1$\times$1$\times$1 convolution helps in reducing the dimensionality and computational complexity. The skip connection facilitates the flow of information, preserving essential features across different layers. The utilization of SimpleGate and SCA enhances the integration of local and global information by suppressing less informative features \cite{zamir2022restormer}.

\begin{figure}[h]
    \centering
    \includegraphics[width=0.48\textwidth]{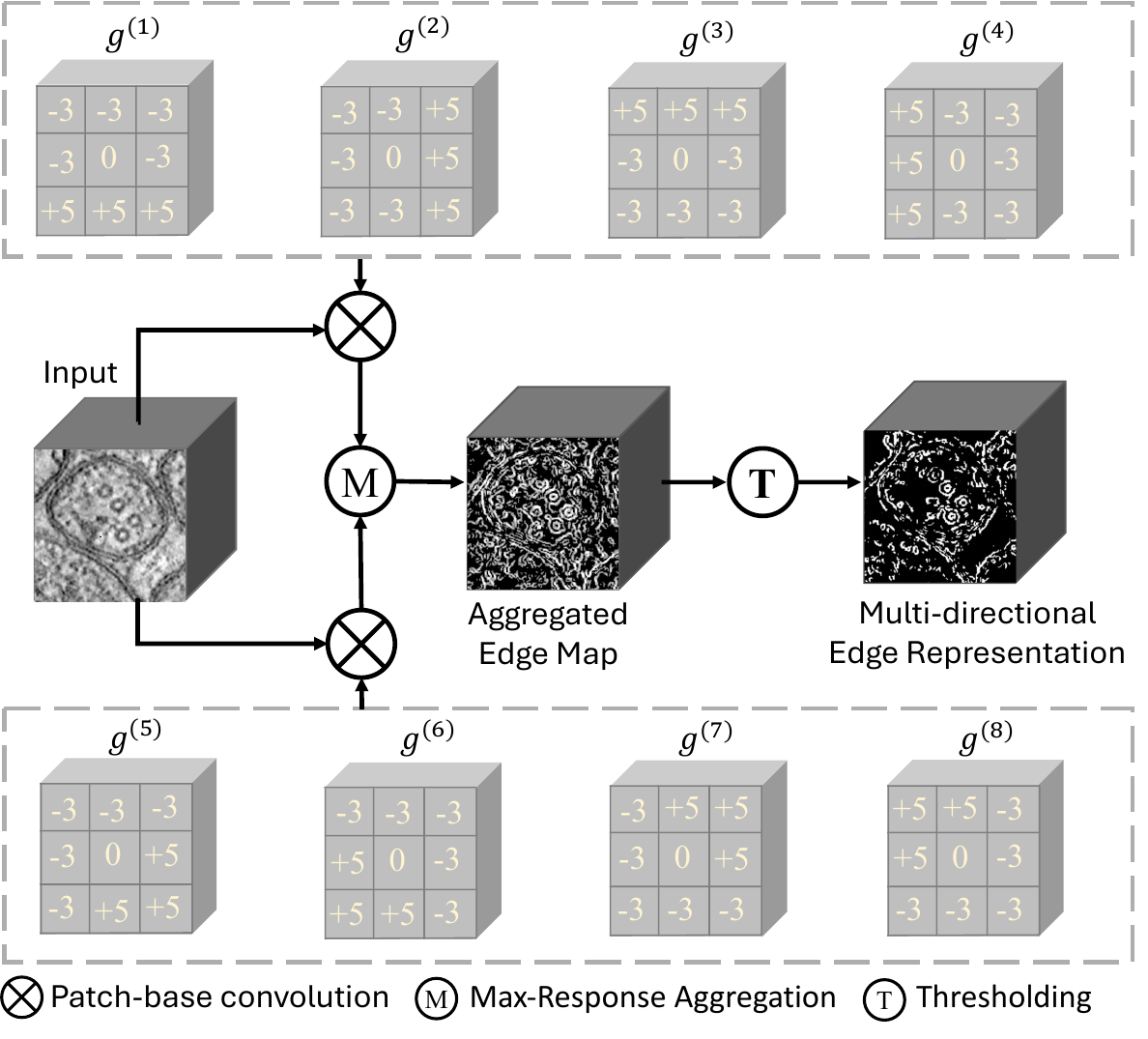} 
    \captionsetup{skip=0pt} %
    \begin{spacing}{0.5}
    \caption{Details of edge representation enhancer.}
    \label{fig:edge_representation}
    \end{spacing}
\end{figure}

\textbf{(4) Edge Representation Enhancer.} 
The edge and texture information captured in the biliteral filtered volume provides complementary guidance for tomographic restoration. As shown in Figure \ref{fig:edge_representation}, this edge representation enhancer excels in extracting detailed edge and contour information by leveraging multi-directional edge detection. This representation significantly aids the training process when integrated into the network. More details are in Supplementary Material S6.

\subsection{Loss Function}

A combined loss function is introduced to our model for both noise removal and structure preservation.

\textbf{Structural reconstruction loss.}
The structural reconstruction loss ensures the denoised output retains the essential details of the original noisy input. Formulated for a BSN, this loss function is defined as:
\vspace{-0.5em}
\begin{equation}
    \mathcal{L}_{\text{rec}} = \mathbb{E} \left[ \left\| f_\theta (V_i(x)) - V_i(x) \right\|_2^2 \right]
    \label{Structural reconstruction loss}
    \vspace{-0.5em}
\end{equation}

where $f_\theta$ is the denoising network parameterized by $\theta$, and $V_i(x)$ represents the $i$-th raw noisy input volumetric patch. By minimizing this loss, the network effectively filters out noise and maintains the integrity of the underlying structures, leading to higher-quality reconstructions and better downstream analysis. 

\textbf{Contrast guidance loss.}
The consrast guidance loss aims to balance noise suppression and the preservation of ultrastructural details. To achieve this, a 3D Gaussian filter is applied to the raw noisy volume, producing a smoothed volume $V_{f}(x)$ that capturing mid- and low-frequency contrast information in order to recover the ultrastructure with higher contrast in the situation of training without ground truth. The loss function is defined as:
\vspace{-0.3em}
\begin{equation}
    \mathcal{L}_{\text{guide}} = \mathbb{E} \left[ \left\| f_\theta (V_{i}(x)) - V_{if}(x) \right\|_2^2 \right]
    \label{Contrast guidance loss}
    \vspace{-0.5em}
\end{equation}
where $V_{if}(x)$ is the $i$-th filtered volumetric patch. 

\textbf{Edge enhancement loss.}
Bilateral filtering can smooth the image while preserving edge information, whereas gaussian filtering tends to blur the edges. Since bilateral filtering takes into account both spatial and intensity domains, it assigns smaller weights to pixels with significant intensity differences, thereby reducing noise while preserving more edges and details. 
We adapt edge enhancement extractor $K$ on network output $f_\theta (V_{i}(x))$ and bilateral filtering reference volume $V_{ib}(x)$ to extract the intensity maps. Thus the edge enhancement loss can be defined as:
\vspace{-0.5em}
\begin{equation}
    \mathcal{L}_{\text{edge}} = \mathbb{E} \left[ \left\| K(f_\theta (V_{i}(x))) - K(V_{ib}(x)) \right\|_2^2 \right]
    \label{Edge enhancement loss}
    \vspace{-0.5em}
\end{equation}

This loss complements the contrast guidance loss by preserving of edges and details that might be lacking. 
\begin{table*}[t]
    \centering
    \captionsetup{skip=0pt} %
   \begin{spacing}{1.2}
    \caption{Quantitative comparison on simulated datasets. (Metrics: PSNR(db)/SSIM). The \textbf{best} results among methods that do not require ground truth are highlighted, and the tilde ($\sim$) indicates approximate running time.}
    \label{Simulated PSNR}
    \end{spacing}
    \resizebox{\linewidth}{!}{
    \begin{tabular}{clcccccccccccccccc}
        \toprule
       \multirow{2}{*}{} & \multirow{2}{*}{Methods} & \multicolumn{3}{c}{Shrec2020} & \multicolumn{3}{c}{Shrec2021} & \multicolumn{3}{c}{PolNet} & Running \\
        &  & 0.1 & 0.15 & 0.2 & 0.1 & 0.15 & 0.2 & 0.1 & 0.15 & 0.2 & time (mins)\\
        \midrule
        & Noisy & 24.14 / 0.598 & 20.63 / 0.440 & 18.12 / 0.377 & 24.36 / 0.252 & 20.84 / 0.179 & 18.34 / 0.142 & 27.19 / 0.601 & 23.67 / 0.448 & 21.17 / 0.344 & \\
        \midrule
        \multirow{6}{*}{Non-learning} &Gaussian & 28.96 / 0.785 & 23.49 / 0.533 & 20.05 / 0.469 & 27.88 / 0.331 & 24.62 / 0.214 & 21.22 / 0.186  & 30.07 / 0.693  & 26.81 / 0.515 & 24.37 / 0.379 & 0.1 \\
        &Bilateral & 26.89 / 0.707 & 22.37 / 0.461 & 18.11 / 0.372 & 27.79 / 0.324 & 22.02 / 0.191 & 18.91 / 0.167 & 29.82 / 0.643 & 25.35 / 0.478 & 24.65 / 0.381 & 0.3\\
        &LPF & 26.34 / 0.673 & 21.41 / 0.449 & 18.96 / 0.386 & 26.15 / 0.285 & 22.74 / 0.198 & 19.19 / 0.153 & 27.56 / 0.610 & 24.14 / 0.453 & 21.49 / 0.352 & 0.5\\
        &NAD & 27.77 / 0.736 & 24.49 / 0.564 & 21.31 / 0.492 & 28.13 / 0.357 & 25.92 / 0.222 & 20.74 / 0.177 & 29.91 / 0.687 & 25.48 / 0.503 & 22.27 / 0.369 & 3\\
        &NLM & 24.96 / 0.612 & 20.34 / 0.437 & 17.79 / 0.351 & 25.09 / 0.267 & 21.12 / 0.188 & 17.98 / 0.137 & 27.82 / 0.613 & 22.75 / 0.424 & 21.83 / 0.351 & 3\\
        &BM4D & 30.11 / 0.868 & 29.33 / 0.862 & 29.73 / 0.877 & 34.77 / 0.817 & 32.59 / 0.779 & 31.81 / 0.736 & 35.26 / 0.893 & 33.36 / 0.849 & 31.54 / 0.814 & 7\\
        \midrule
        \multirow{3}{*}{Supervised}&DnCNN & 39.67 / 0.966 & 39.57 / 0.959 & 38.92 / 0.954 & 42.77 / 0.944 & 42.54 / 0.947 & 41.85 / 0.938 & 42.36 / 0.985 & 42.53 / 0.979 & 40.82 / 0.971 & $\sim$ 13\\
        &FFDNet & 41.53 / 0.971 & 41.51 / 0.976 & 40.89 / 0.963 & 43.06 / 0.952 & 42.19 / 0.942 & 42. 26 / 0.931 & 42.73 / 0.981 & 41.77 / 0.968 & 41.13 / 0.961 & $\sim$ 12\\
        &V-Net & 40.82 / 0.967 & 40.14 / 0.966 & 39.92 / 0.952 & 42.81 / 0.961 & 41.11 / 0.957 & 40.95 / 0.946 & 42.48 / 0.987 & 40.54 / 0.972 & 40.89 / 0.953 & $\sim$ 15\\ 

        \midrule
        \multirow{4}{*}{Self-supervised}&N2V & 24.38 / 0.602 & 20.19 / 0.428 & 17.55 / 0.363 & 24.13 / 0.241 & 20.69 / 0.168 & 18.29 / 0.143 & 26.54 / 0.592 & 22.93 / 0.426 & 21.21 / 0.341 & $\sim$ 20\\
        &SC-Net & 31.22 / 0.921 & 30.01 / 0.903 & 28.44 / 0.878 & 33.96 / 0.796 & 30.15 / 0.771 & 29.78 / 0.704& 32.73 / 0.927 & 30.51 / 0.895 & 27.62 / 0.852 & $\sim$ 25\\
        &NMSG & \textbf{36.10} / \textbf{0.944} & 34.83 / 0.923 & 34.91 / 0.915 & 36.47 / 0.812 & 35.56 / 0.809  & 35.17 / 0.766 & 38.28 / \textbf{0.975} & 36.44 / 0.929 & 34.07 / 0.893 & $\sim$ 50\\
    
        & \textbf{Ours} & 36.02 / 0.941 & \textbf{36.18} / \textbf{0.952} & \textbf{35.79} / \textbf{0.936} & \textbf{39.23} / \textbf{0.832} & \textbf{38.16} / \textbf{0.810} & \textbf{37.64} / \textbf{0.784} & \textbf{38.90} / 0.966 & \textbf{37.06} / \textbf{0.953} & \textbf{36.51} / \textbf{0.932} & $\sim$ 35\\
        \bottomrule
    \end{tabular}
    }
    \vspace{-1em}
\end{table*}

\textbf{Total variation loss.} Total variation loss (TV Loss) is a regularization method commonly used in image denoising and image reconstruction. It works by minimizing the total variation of the image gradient, which smooths local regions without over-blurring edges. The TV loss for a volumetric image is defined as:
\small
\vspace{-0.5em}
\begin{equation}
    \mathcal{L}_{TV} = \sum_{r,s,t} \left(|\nabla_x V(r,s,t)| + |\nabla_y V(r,s,t)| + |\nabla_z V(r,s,t)|\right)
    \label{TV Loss}
\end{equation}
\normalsize
where $V(r,s,t)$ represents the voxel value at position $(r,s,t)$ in the network output $f_\theta (V_{i}(x))$. $\nabla_x, \nabla_y, \nabla_z$ represent the gradients in the $x, y$, and $z$ directions, respectively. By penalizing large gradients, TV loss maintains the structural integrity of the image while reducing noise.

Eq.~\ref{Structural reconstruction loss}, \ref{Contrast guidance loss}, \ref{Edge enhancement loss}, \ref{TV Loss} are combined to formulated the complete loss function of the denoising network:
\vspace{-0.5em}
        \begin{equation}
            \mathcal{L}_{\text{total}} = \lambda_1 \mathcal{L}_{\text{rec}} + \lambda_2 \mathcal{L}_{\text{guide}} + \lambda_3 \mathcal{L}_{\text{edge}} +
        \lambda_4 \mathcal{L}_{\text{TV}}
        \label{Eq: combined loss}
        \vspace{-0.5em}
        \end{equation}
where $\lambda_1, \lambda_2$, $\lambda_3$ and $\lambda_4$ are the weight parameters of the loss functions.

\section{Experiments}
We evaluated the performance of our method on three simulated datasets and four real Cryo-ET datasets, The proposed model is compared with single volume-based methods including non-learning methods Gaussian filter, Bilateral filter \cite{tomasi1998bilateral}, Low-pass filter (LPF), NAD \cite{frangakis2001noise}, NLM filter \cite{buades2005non}, BM4D \cite{maggioni2012nonlocal}. We also compare it with supervised method like DnCNN \cite{zhang2017beyond}, FFDNet \cite{zhang2018ffdnet}, and V-Net \cite{milletari2016v}, and self supervised learning based method like Noise2Void (N2V) \cite{krull2019noise2void}, SC-Net \cite{yang2021self} and NMSG \cite{yang2024self}, resulting in a detailed quantitative empirical analysis. The sources of datasets can be found in Supplementary Material S3.

\subsection{Networks Training Details}
\begin{figure*}[h!]
    \centering
    \includegraphics[width=0.9\textwidth]{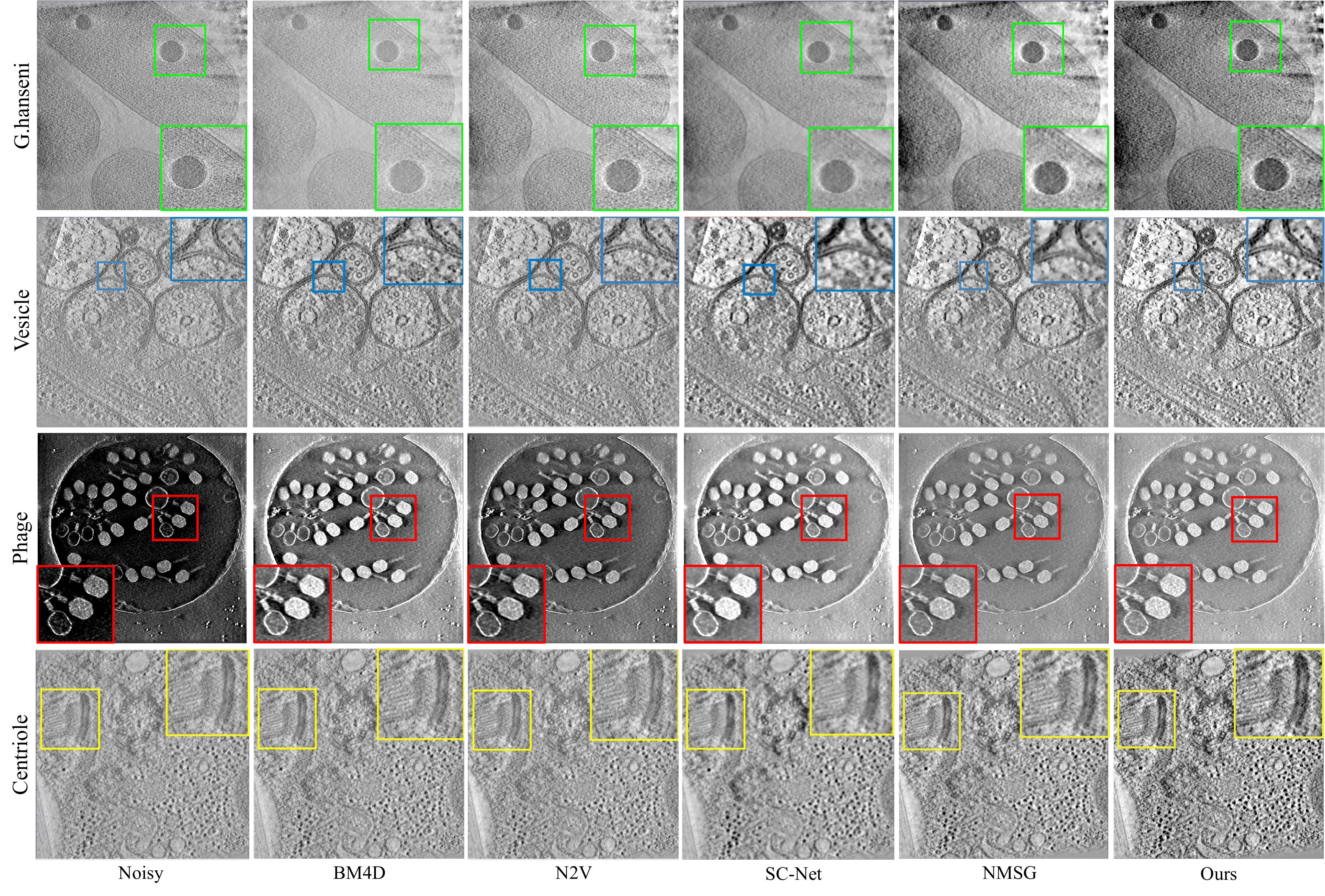}
    \captionsetup{skip=0pt} %
    \begin{spacing}{0.5}
    \caption{Visual results of the real data.}
     \label{fig: Experiments on Real datasets}
    \end{spacing}
    \vspace{-1em}
\end{figure*}

Our method is implemented by Pytorch \cite{paszke2019pytorch} with the model was trained on one NVDIA GeForce RTX 4090 GPU for all the experiments. The model is trained with single noisy volume each time. During the training, the batch size was set to 2 with $108^3$ patch size and 20$\%$ of the patches are selected as validation. The model was trained by 15 epochs for each noisy data, where the optimizer is Adam \cite{diederik2014adam} with $\beta_1=0.5$ and $\beta_2=0.999$. The learning rate was set to 0.0002. For Eq.~\ref{Eq: combined loss}, the parameters were set as $\lambda_1=0.8, \lambda_2=0.5$, $\lambda_3=0.05$ and $\lambda_4=0.01$. 
\subsection{Experiments on Simulated Data}
To assess the performance of our proposed method, we conducted experiments on three simulated datasets: SHREC2020, SHREC2021, and a private dataset obtained using the simulation software PolNet \cite{martinez2023simulating}. These datasets were generated with varying levels of additive white Gaussian noise (AWGN) at intensities $\sigma$ = 0.1, 0.15 and 0.2, following the procedure outlined in \cite{yang2021self}.

The PSNR and SSIM values in Table \ref{Simulated PSNR} demonstrate the robustness of our method in maintaining structural integrity while effectively reducing noise. Supervised methods in Table \ref{Simulated PSNR} tend to have faster running times, but their dependence on clean data limits their real-world applicability in Cryo-ET. Despite the inherent limitation of not using pixel information directly and slightly slower, our method demonstrates superior performance among self-supervised approaches, offering a balanced trade-off between denoising quality and applicability in scenarios without paired datasets. Visual comparison can be found in Supplementary Material S7.

\begin{table}[h]
\centering
\captionsetup{skip=0pt} %
\begin{spacing}{1.2}
\caption{Quantitative results for different missing wedge levels. (AWGN:$\sigma$=0.2, metrics: PSNR(db)/SSIM)}
\label{Tab: missing wedge effect}
\end{spacing}
\resizebox{\linewidth}{!}{
\begin{tabular}{cccc}
\toprule
 Angular Range & MW Level & Noisy &  Denoised \\
\midrule
 $-70.0^\circ$ to  $+70.0^\circ$ (step $2^\circ$) & Moderate & 23.72 / 0.496 & \textbf{38.53} / \textbf{0.948} \\
 $-60.0^\circ$ to  $+60.0^\circ$ (step $2^\circ$)& Standard & 21.59 / 0.371 & 36.48 / 0.934\\
 $-40.0^\circ$ to  $+40.0^\circ$ (step $2^\circ$)& Severe & 16.81 / 0.304 & 35.49 / 0.906\\
 $-30.0^\circ$ to  $+30.0^\circ$ (step $2^\circ$)& Extreme & 12.63/ 0.234 & 31.92 / 0.822\\

\bottomrule
\end{tabular}
}
\vspace{-0.5em}
\end{table}
To assess the missing wedge (MW) effect \cite{frank1996spider} on our model, we generated simulated tomograms with varying MW levels by adjusting the tilt series angular range. As shown in Table \ref{Tab: missing wedge effect}, despite performance decreases with more severe MW, the model still preserves structure well, demonstrating its robustness.

\subsection{Real-world Datasets}

We evaluate on four real cryo-ET datasets: G. hansenii bio9-2, Vesicle, Escherichia phage T4, and Centriole. The G. hansenii bio9-2 dataset is a tilt series of 61 projections ranging from  $-60^\circ$ to $+60^\circ$ at $2^\circ$ intervals. Each tilt image measures $ 960 \times 928$ pixels with 5.41 \AA/pixel. The Vesicle dataset consists of 120 projections ranging from $-59^\circ$ to $+60^\circ$ at $1^\circ$ intervals. Each tilt image measures 1024 $\times$ 1024 pixels with 8 \AA/pixel. The Escherichia phage T4 dataset contains 41 projections ranging from $-60^\circ$ to $+60^\circ$ at $3^\circ$ intervals. Each tilt image measures 1024 $\times$ 1024 pixels with 1.558 \AA/pixel. The Centriole dataset is a tilt series of 64 projections ranging from $-61.0^\circ$ to $+65.0^\circ$ at $2^\circ$ at intervals. Each tilt image measures 1024 $\times$ 1024 pixels with 10.1 \AA/pixel. The tomographic volumes for these specimens were reconstructed using the \textit{tilt} program in IMOD\cite{kremer1996computer}. 

\subsection{Experiments on Real Cryo-ET data}
\begin{figure}[ht]
    \centering
    \includegraphics[width=\linewidth]{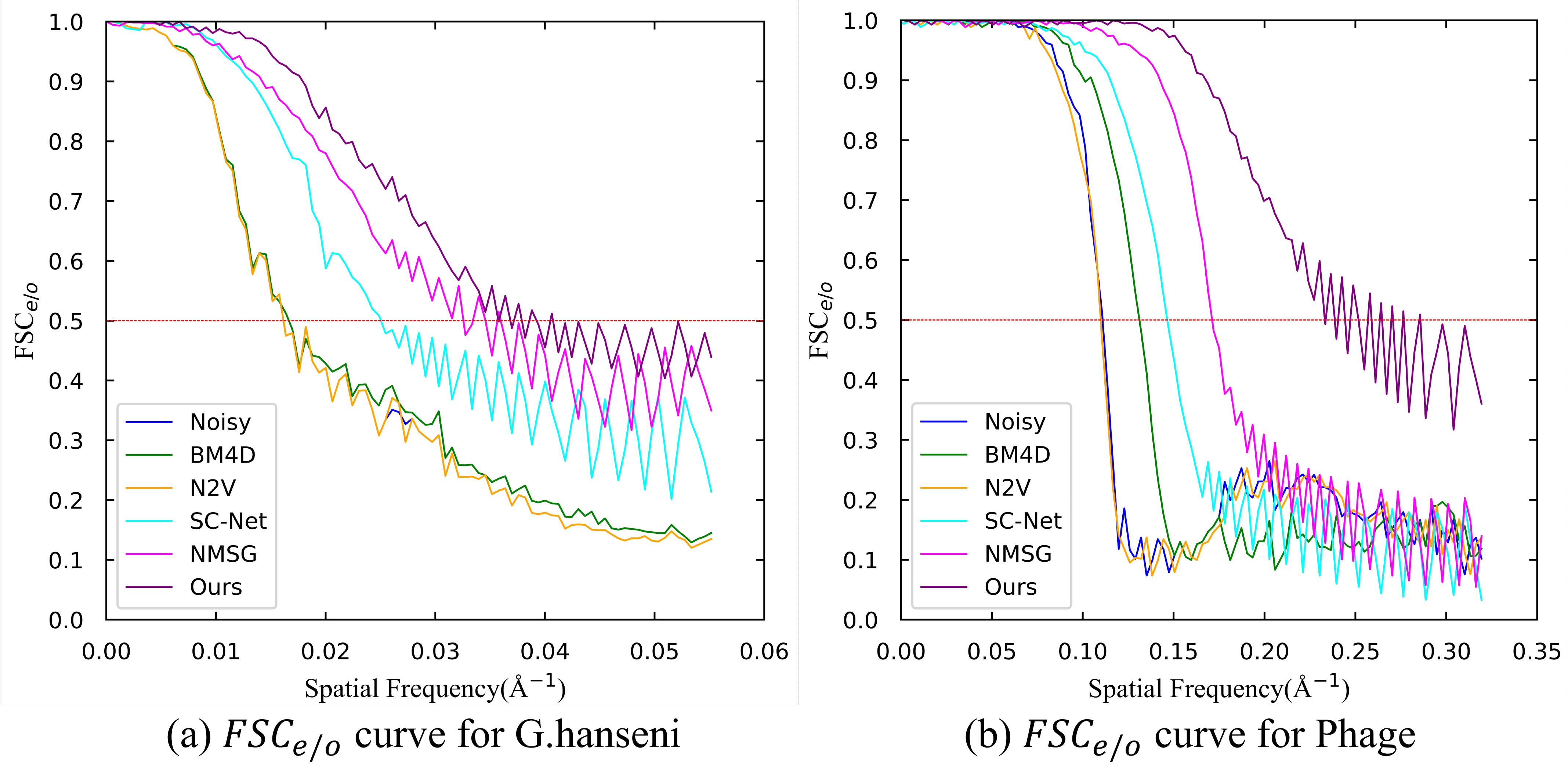}
    \captionsetup{skip=0pt} %
    \begin{spacing}{0.5}
    \caption{Examples of FSC$_{e/o}$ curves for the G.hanseni and Phage. Red dash line in the figures point out the position of FSC$_{e/o}$=0.5.}
    \label{fig:fsc curve}
    \end{spacing}
    \vspace{-0.8em}
\end{figure}

\textbf{Visual analysis.} Figure \ref{fig: Experiments on Real datasets} shows the visualized results of the real world cryo-ET datasets. From the results of G.hanseni, we can clearly see that the volume denoised by the our method are recovered with the best contrast. The phage structures in the Escherichia phage T4 can be easily identified without the darkness and edge blurness after denoising, while others detailed structures remain blurred and dark. The datasets Vesicle and Centriole contain complex cellular structures which are difficult to denoise without abundant pre-defined knowledge. In such cases, our model can still derive denoised volumes with enough contrast, while preserving the ultrastructure integrity. The denoised volumes derived by the proposed model introduce fewer grid artifacts than the other results. Figure \ref{fig:fsc curve} presents FSC$_{e/o}^{-1}(0.5)$ curves for the G.hanseni and Phage, which illustrates that our method can perform better image restoration with higher self-consistency and introduce fewer false signal comparing with other methods.
\vspace{-1.0em}

\textbf{Quantitative analysis.} As ground-truth is unavailable in real-world situations, we adopt a cross-validation metric called FSC\(_{e/o}\) \cite{cardone2005resolution} to assess the resolution of cryo-ET volumes. Table \ref{Tab: Real FSC Resolition} shows the FSC$_{e/o}^{-1}(0.5)$ resolution calculated on the four real-world data. All the tomogram is calculated with subtomogram sized by $512^2\times100$. Results show that our method achieves the best resolution among other volumetric image denoising methods.
\vspace{-0.5em}
\begin{table}[h]
\centering
\captionsetup{skip=0pt} %
\begin{spacing}{1.2}
\caption{Resolution estimated by FSC$_{e/o}^{-1}(0.5)$ (Angstrom, \AA). For resolution of a tomogram, the lower is the better.}
\label{Tab: Real FSC Resolition}
\end{spacing}
\begin{tabular}{ccccc}
\toprule
Dataset & G.hanseni &  Vesicle &  Phage &  Centriole \\
\midrule
Noisy & 63.43 & 30.05 & 10.17 & 57.74 \\
BM4D & 61.08 & 29.44 & 9.30 &  57.37\\
N2V & 63.41 &  30.01 & 10.15 &  57.65\\
SC-Net & 40.22 &  22.67 & 6.93 & 38.21 \\
NMSG & 27.95 & 20.11 & 5.92 &  31.10\\
Ours & \textbf{24.61} & \textbf{17.38} & \textbf{3.50} &  \textbf{26.96}\\
\bottomrule
\end{tabular}
\vspace{-1.4em}
\end{table}

\section{Ablation studies}

\begin{figure}[ht]
    \centering
    \includegraphics[width=\linewidth]{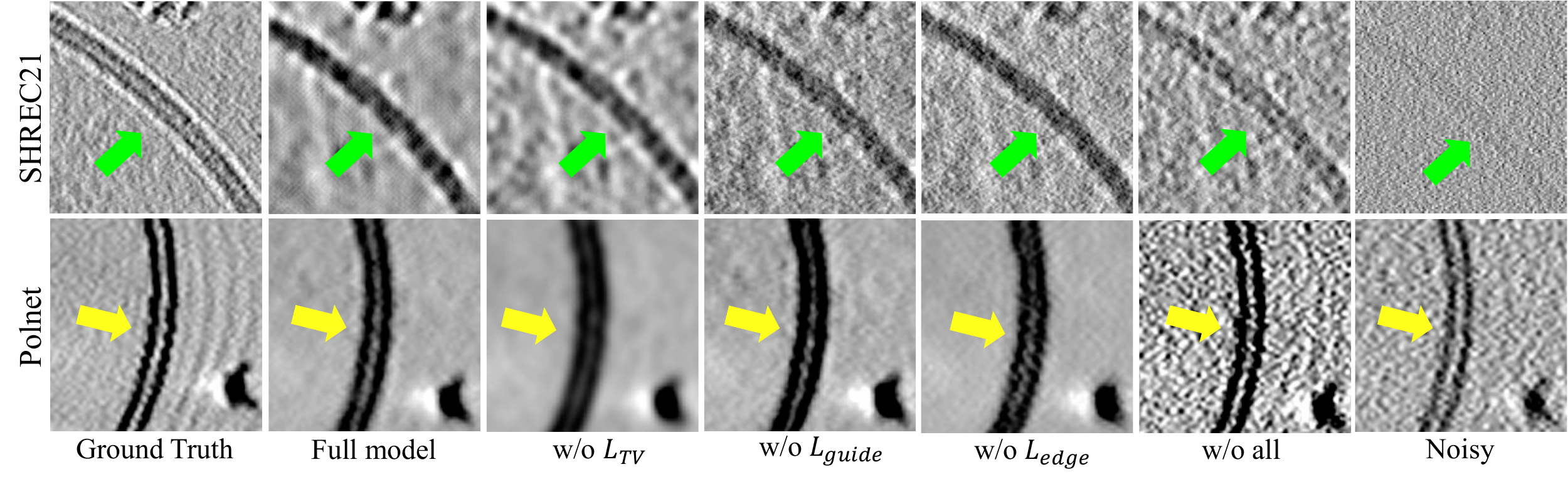}{}
    \captionsetup{skip=0pt} %
    \begin{spacing}{0.5}
    \caption{Visual results of component analysis on loss functions.}
    \label{fig:Ablation Study 1}
    \end{spacing}
    \vspace{-0.8em}
\end{figure}

\paragraph{Component analysis of loss function.} To verify the effectiveness of our combined loss function, we conducted component analysis by systematically removing one or more components ($\mathcal{L}_{guide}$, $\mathcal{L}_{edge}$ and $\mathcal{L}_{TV}$) and comparing the results with the full model. Table \ref{Tab: Ablation Study 1} indicates that removing any component significantly reduces performance, highlighting their importance in guiding the network to preserve contrast, edge, and texture information while regularizing the model. The full model produces the highest quality results (Figure \ref{fig:Ablation Study 1}), with better preservation of structural details and fewer artifacts. Thus, the combined loss function ensures balanced noise suppression and structural preservation, leading to high-quality denoised volumetric images.

\begin{table}[ht]
\centering
\captionsetup{skip=0pt} %
\begin{spacing}{1.2}
\caption{PSNR/SSIM results for ablation study on the component analysis of the loss functions. (AWGN:$\sigma$=0.2).}
\label{Tab: Ablation Study 1}
\end{spacing}
\begin{tabular}{ccc}
\toprule
Method & SHREC21 & PolNet \\
\midrule
Ours (w/o $\mathcal{L}_{edge}$) & 30.71 / 0.695 & 31.29 / 0.819  \\
Ours (w/o $\mathcal{L}_{guide}$) & 35.93 / 0.644& 33.24 / 0.875 \\
Ours (w/o $\mathcal{L}_{TV}$) & 36.94 / 0.736 & \textbf{36.89} / 0.917  \\
Ours (w/o all) & 28.99 / 0.431 & 26.64 / 0.507 \\
Ours (full model) & \textbf{37.62} / \textbf{0.784}& 36.51 /  \textbf{0.932} \\
\bottomrule
\end{tabular}
\vspace{-1.6em}
\end{table}

\begin{figure}[ht]
    \centering
    \includegraphics[width=\linewidth]{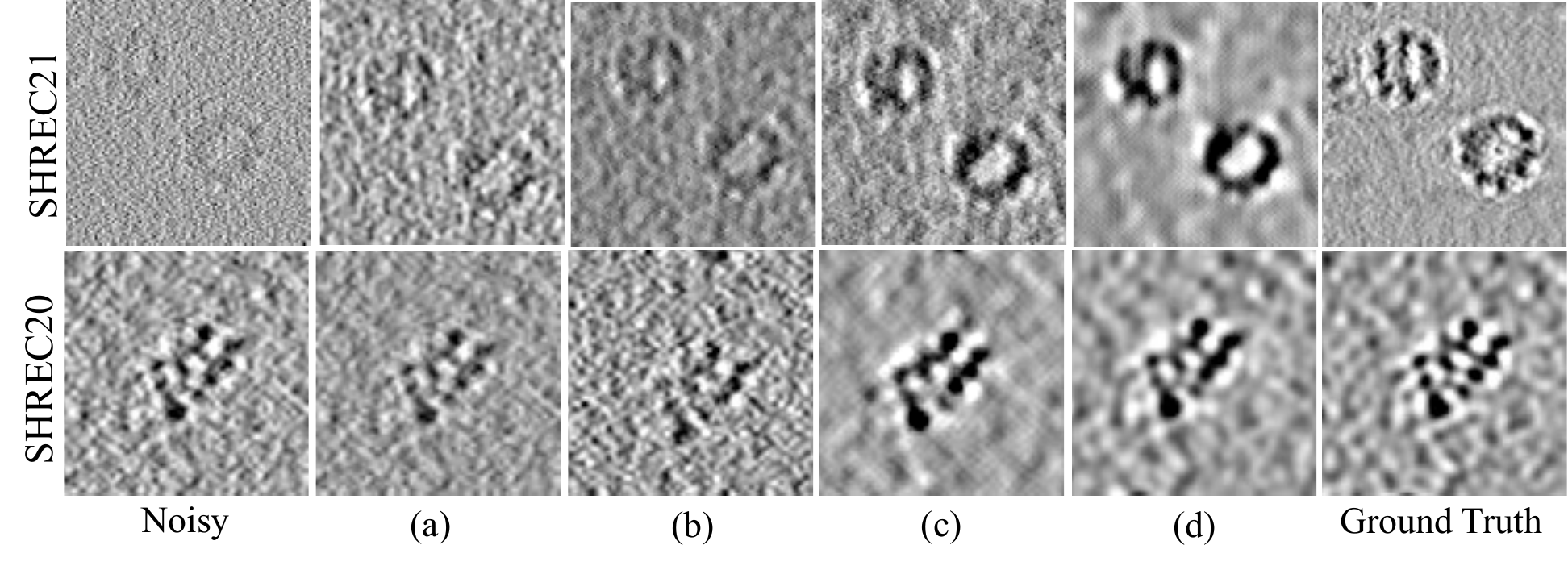}{}
    \captionsetup{skip=0pt} %
    \begin{spacing}{0.5}
    \caption{Visual results of study on volume unsuffle/shuffle.}
    \label{fig:Ablation Study 3}
    \end{spacing}
    \vspace{-1.0em}
\end{figure}

\paragraph{Study on noise patterns in simulated data.} To rigorously assess the model's robustness across various noise conditions, we generated simulated tomograms using different noise patterns. Specifically, we introduced Gaussian noise ($\sigma$=0.2) to represent a common baseline scenario and Poisson noise ($\lambda$=0.02). We also included a mixture noise model with respective weights adjusted to ensure an overall noise level comparable to the individual noise types.
\vspace{-0.5em}
\begin{table}[ht]
    \centering
    \captionsetup{skip=0pt} %
   \begin{spacing}{1.2}
    \caption{Quantitative result of proposed methods under different noise patterns on simulated datasets. For each column, the \textbf{best} and \underline{second-best} values are highlighted. (metrics: PSNR/SSIM) }
    \end{spacing}
    \resizebox{\linewidth}{!}{
    \begin{tabular}{ccccccccccc}
        \toprule
        Noise &  &\multicolumn{1}{c}{Shrec2020} & \multicolumn{1}{c}{Shrec2021} & \multicolumn{1}{c}{PolNet} \\

        \midrule
        Gaussian & noisy & 18.12 / 0.377  & 18.34 / 0.142  & 21.17 / 0.344 \\
        ($\sigma=0.2$) & denoised  & \textbf{35.79} / \textbf{0.936} & \textbf{37.64} / \textbf{0.784}  & \textbf{36.51} / \textbf{0.932} \\
        \midrule
        Poisson & noisy  & 16.81 / 0.332 & 17.23 / 0.129 & 17.56 / 0.306 \\
        ($\lambda$=0.02) & denoised & \underline{31.57} / \underline{0.873} & \underline{31.49} / \underline{0.695} & \underline{33.32} / \underline{0.901} \\
        \midrule
        Mixture & noisy  & 16.11 / 0.293 & 16.59 / 0.104 & 17.44 / 0.273 \\
        (weights adjusted)& denoised & 25.96 / 0.522 & 24.81 / 0.371 & 26.42 / 0.686\\
       
        \bottomrule
    \end{tabular}
    }

    \vspace{-1em}
\end{table}
\vspace{-1.2em}
\paragraph{Study on network components.}
The effect of the DCA module and volume unshuffle/shuffle (VU/S) is illustrated in Figure \ref{fig:Ablation Study 3} and detailed in Table \ref{Tab: Ablation Study 3}. In Table \ref{Tab: Ablation Study 3} (a), we adopt a standard 3-layer UNet. For Table \ref{Tab: Ablation Study 3} (b), we use max pooling for downsampling and bilinear interpolation for upsampling. For Table \ref{Tab: Ablation Study 3} (c), the D-BSN block replaces replaces the DCA module. Overall, the combination of DCA and VU/S (Table \ref{Tab: Ablation Study 3} (d)) exhibits the best performance compared to other ablated results.

\begin{table}[h]
\centering
\captionsetup{skip=0pt} %
\begin{spacing}{1.2}
\caption{Quantitative analysis of the effects of the DCA module and VU/S. (AWGN:$\sigma$=0.10, metrics: PSNR(db)/SSIM).}
\label{Tab: Ablation Study 3}
\end{spacing}
\resizebox{\linewidth}{!}{
\begin{tabular}{lcccc}
\toprule
& DCA & VU/S & SHREC20 &  SHREC21 \\
\midrule
(a) UNet & - & - & 25.62 / 0.593 & 24.33 / 0.249 \\
(b) DCA+UNet & $\checkmark$ & - & 26.25 / 0.607 & 23.93 / 0.255\\
(c) D-BSN+VU/S  & - & $\checkmark$ & 31.81 / 0.893 & 33.79 / 0.726\\
(d) DCA+VU/S & $\checkmark$ & $\checkmark$ & \textbf{36.02} / \textbf{0.941} & \textbf{39.23} / \textbf{0.832}\\

\bottomrule
\end{tabular}
}
\end{table}

\vspace{-1.6em}
\section{Conclusions}
This study presents a novel U-shape self-supervised learning model for denoising Cryo-ET volumetric images. By incorporating a $\mathcal{J}$-invariant BSN and volume-unshuffle/shuffle technique, our model demonstrates superior noise reduction and structural preservation compared to existing methods. The comprehensive results validate the effectiveness of our approach, offering a powerful tool for researchers in structural biology to achieve more accurate and detailed visualizations of cellular structures.

{\small
\bibliographystyle{ieee_fullname}
\bibliography{egbib}
}

\end{document}